\documentclass[reprint,superscriptaddress,aps,prb,twocolumn,floatfix]{revtex4-2}
\usepackage{setspace}
\usepackage{amsmath}
\usepackage{bm}
\usepackage{graphicx}
\usepackage[nearskip,margin = 0pt]{subfig}
\usepackage{verbatim}
\usepackage{amsfonts}
\usepackage{amssymb}
\usepackage{textcomp}
\usepackage{mathrsfs}
\usepackage{mathtools}
\usepackage{url}
\usepackage{caption}
\usepackage{upgreek}

\usepackage{xcolor}
\usepackage[colorlinks,linkcolor=blue,anchorcolor=blue,citecolor=blue,urlcolor=black]{hyperref}
\usepackage{ragged2e}
\usepackage{float}
\usepackage{lineno}
\DeclareGraphicsExtensions{.pdf,.eps,.png,.jpg,.mps}
\begin{document}

\title{
Highly coherent two-color laser with stability below $\mathbf{3\times 10^{-17}}$ at 1 second
}
\author{Bibo He$^{1,\star}$, Jiachuan Yang$^{1,\star}$, Fei Meng$^{2,\star,\dagger}$, Jialiang Yu$^{3}$, Chenbo Zhang$^{1}$, Qi-Fan Yang$^{4}$,\\ Yani Zuo$^{2}$, Yige Lin$^{2}$, Zhangyuan Chen$^{1}$, Zhanjun Fang$^{2}$, Xiaopeng Xie$^{1,\ddagger}$\\
\vspace{3pt}
$^1$State Key Laboratory of Advanced Optical Communication Systems and Networks,\\School of Electronics, Peking University, Beijing 100871, China\\
$^2$Division of Time and Frequency Metrology, National Institute of Metrology, Beijing 100029, China\\
$^3$Physikalisch-Technische Bundesanstalt, Bundesallee 100, 38116 Braunschweig, Germany\\
$^4$State Key Laboratory for Artificial Microstructure and Mesoscopic Physics and Frontiers Science Center for Nano-optoelectronics, School of Physics, Peking University, Beijing, 100871, China\\
$^\star$These authors contributed equally to this work\\
\vspace{3pt}
Corresponding authors: $^\dagger$mfei@nim.ac.cn, $^\ddagger$xiaopeng.xie@pku.edu.cn.}



\maketitle
\noindent
\large\textbf{Abstract} \\
\normalsize\textbf{
Two-color lasers with high coherence are paramount in precision measurement, accurate light-matter interaction, and low-noise photonic microwave generation. However, conventional two-color lasers often suffer from low coherence, particularly when these two colors face large frequency spacings. Here, harnessing the Pound-Drever-Hall technique, we synchronize two lasers to a shared ultra-stable optical reference cavity to break through the thermal noise constraint, achieving a highly coherent two-color laser. With conquering these non-common mode noises, we demonstrate an exceptional fractional frequency instability of $\mathbf{2.7\times 10^{-17}}$ at 1 second when normalized to the optical frequency. Characterizing coherence across large frequency spacings poses a significant challenge. To tackle this, we employ electro-optical frequency division to transfer the relative stability of a 0.5 THz spacing two-color laser to a 25 GHz microwave signal. As its performance surpasses the sensitivity of the current apparatus, we establish two independent systems for comparative analyses. The resulting 25 GHz signals exhibit exceptional phase noise of $-$74 $\mathbf {dBc\ Hz^{-1}}$ at 1 Hz and $-$120 $\mathbf {dBc\ Hz^{-1}}$ at 100 Hz, demonstrating the two-color laser’s performance approaching the quantum noise limit of its synchronization system. It also sets a new record for the two-point frequency division method in photonic microwave generation. Our achievement in highly coherent two-color lasers and low-noise microwave signals will usher in a new era for precision measurements and refine the accuracy of light-matter and microwave-matter interactions to their next decimal place.
}

\begin{figure*}[ht]
\centering
\captionsetup{singlelinecheck=no, justification = RaggedRight}
\includegraphics[width=17.4cm]{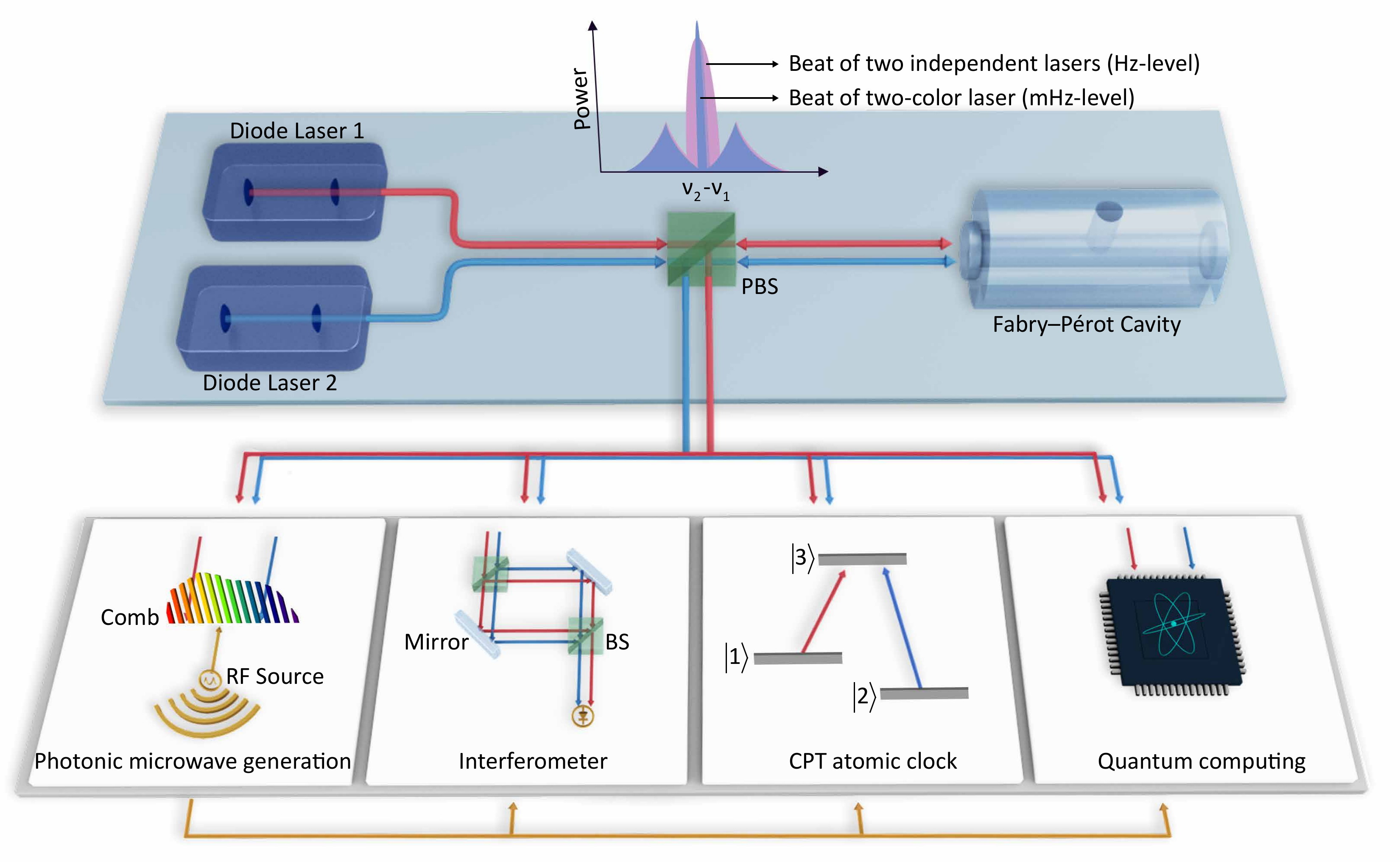}
\caption{\textbf{Concept and architecture of the two-color laser and its applications.} Utilizing the PDH method to synchronize two diode lasers to a shared reference cavity can create a highly coherent two-color laser. Consequently, the relative linewidth of a two-color laser is notably narrower than that of a one-color laser. This highly coherent two-color laser is expected to find applications in photonic microwave generation, precision measurement, and light-matter interaction. In photonic microwave generation, the exceptional coherence of the two-color laser can be downconverted to produce a high-purity microwave signal. Employing this highly coherent two-color laser as the seed source in an interferometer, such as a dual-comb system, can significantly enhance measurement precision. Two-color lasers also prove valuable in light-matter interactions, encompassing applications like CPT atomic clocks, quantum computing, and electromagnetically induced transparency, enhancing overall performance. Furthermore, the pristine microwave signals generated by the highly coherent two-color lasers can be directly utilized in interferometers and microwave-matter interactions, enhancing applications such as very long baseline interferometry, radar, navigation, CPT atomic clocks, and quantum computing.
}
\label{fig1}
\end{figure*}

\vspace{3pt}
\maketitle
\noindent
\large\textbf{Introduction} \\
\normalsize
\noindent 
It has never stopped striving to control the accuracy of optical phase. Whether in precision interferometry or light-matter interactions, advancements in optical phase precision will pave the way for a fresh perspective on our understanding of the physical world. Manipulating the relative phase stability (coherence) of two-color waves is often more feasible than achieving absolute phase accuracy for each color. Given this reality, highly coherent two-color lasers find significant applications in precision interferometric measurements \cite{1gravitational,2gravitational,interferometry}, dual-comb spectroscopy \cite{Dual-comb,tsao2024dual,picque2019frequency}, and play a crucial role in highly accurate optical-atomic interactions, including coherent population trapping (CPT) atomic clocks \cite{2CPT}, electromagnetically induced transparency \cite{EIT}, and quantum computing \cite{quantum}, as shown in Fig. \ref{fig1}. Recent advancements have demonstrated their key role in low-noise photonic terahertz and microwave generation \cite{OFD_LJ, LJoptica,nature1,nature2,nature3, YQF, Vahala, LJarx,kudelin2024tunable,zhang2024monolithic,liu2024low,kittlaus2021low, Brillouin}, where there is a growing demand for two-color lasers with terahertz frequency spacings. Traditional methods of generating two-color lasers, such as two-mode lasers \cite{two-mode} and electro-optic (EO) or acoustic-optic modulation \cite{EOmod}, often encounter challenges related to poor coherence between the two colors. This coherence issue exacerbates as the frequency spacing between the colors widens. Moreover, the modulator approach restricts the achievable frequency spacing. Hence, the urgency to achieve a two-color laser with an extended frequency spacing and high coherence is more pronounced than ever.


Cutting-edge one-color lasers with exceptional coherence and linewidth at the millihertz level are achievable through the Pound-Drever-Hall (PDH) technique \cite{PDH_Hall,40mHz,mHz}. In this method, a diode laser is precisely locked to an optical reference cavity characterized by remarkably high finesse and isolated within a vacuum chamber. These high-performance lasers not only support the most precise atomic clocks \cite{clock} but also enable the generation of pristine microwave signals \cite{xie, 2011OFD, NISTscience}, playing a crucial role in gravitational wave detection \cite{1gravitational, 2gravitational}. The coherence of these PDH-locked lasers is primarily limited by the thermal noise of the optical reference cavity, while the electronic noise associated with the PDH locking mechanism remains significantly lower \cite{yujialiang}. Drawing inspiration from this, synchronizing two diode lasers to a shared optical reference cavity using the PDH technique could enhance the coherence between the two colors to a level constrained only by the electronic noise of each PDH locking system \cite{Hall, chenqunfeng}. In this way, the coherence level of the two-color laser would markedly exceed that of each one-color laser. It opens up new possibilities for achieving two-color lasers with exceptionally high coherence even across wider frequency spacings.

In this work, we employ the PDH technique to synchronize two independent lasers to a shared cavity as illustrated in Fig. \ref{fig1}. Initially, we undertake a thorough analysis to discern the impact of thermal noise and electronic noise on the coherence performance of the two-color laser. On top of that, we establish a two-color laser with a 1.5 GHz spacing by synchronizing two diode lasers with a pair of adjacent longitudinal modes of a shared optical reference cavity. After overcoming these non-ideal phase modulation noises and other non-common mode noises, the relative phase noise of the two-color laser reaches $-$52 $\rm {dBc\ Hz^{-1}}$ at 1 Hz, which is four orders of magnitude lower than that of each one-color laser, and $-$92 $\rm {dBc\ Hz^{-1}}$ at 100 Hz. This phase noise performance is primarily limited by the electronic noise of the PDH locking mechanism and is approaching the state-of-the-art level \cite{yujialiang}, indicating the high coherence of this two-color laser. The phase noise of the two-color laser is expected to remain the same value when gradually increasing the frequency spacing of these two colors. However, assessing this particular property poses challenges. To tackle this, we employ an EO comb-based optical frequency division system to transfer the relative stability of the two-color laser to a microwave frequency signal for characterization \cite{OFD_LJ}. Yet, the phase noise of the down-converted microwave signal surpasses the sensitivity of conventional phase noise analyzers, prompting us to establish two identical and independent systems. Specifically, we configure the two-color laser with a 0.5 THz spacing, transferring its relative stability to a 25 GHz microwave signal. Two identical systems generate two microwave signals, with their beat note reflecting their phase noise, enabling us to reference the phase noise of the two-color lasers. The phase noise of the down-converted 25 GHz signal reaches $-$74 dBc/Hz at 1 Hz and $-$120 dBc/Hz at 100 Hz, affirming the high coherence of the two-color laser even with a significant frequency spacing. Notably, the EO comb-based optical frequency division system not only provides a setup for laser phase noise characterization but also represents a robust, stable, and low phase noise configuration for microwave generation, heralding a new era in this field. Our simple and efficient structure lays the groundwork for highly coherent two-color laser and low phase noise microwave applications, making them more accessible and practical beyond traditional laboratory environments.

\begin{figure*}[ht]
\centering
\captionsetup{singlelinecheck=no, justification = RaggedRight}
\includegraphics[width=17.4cm]{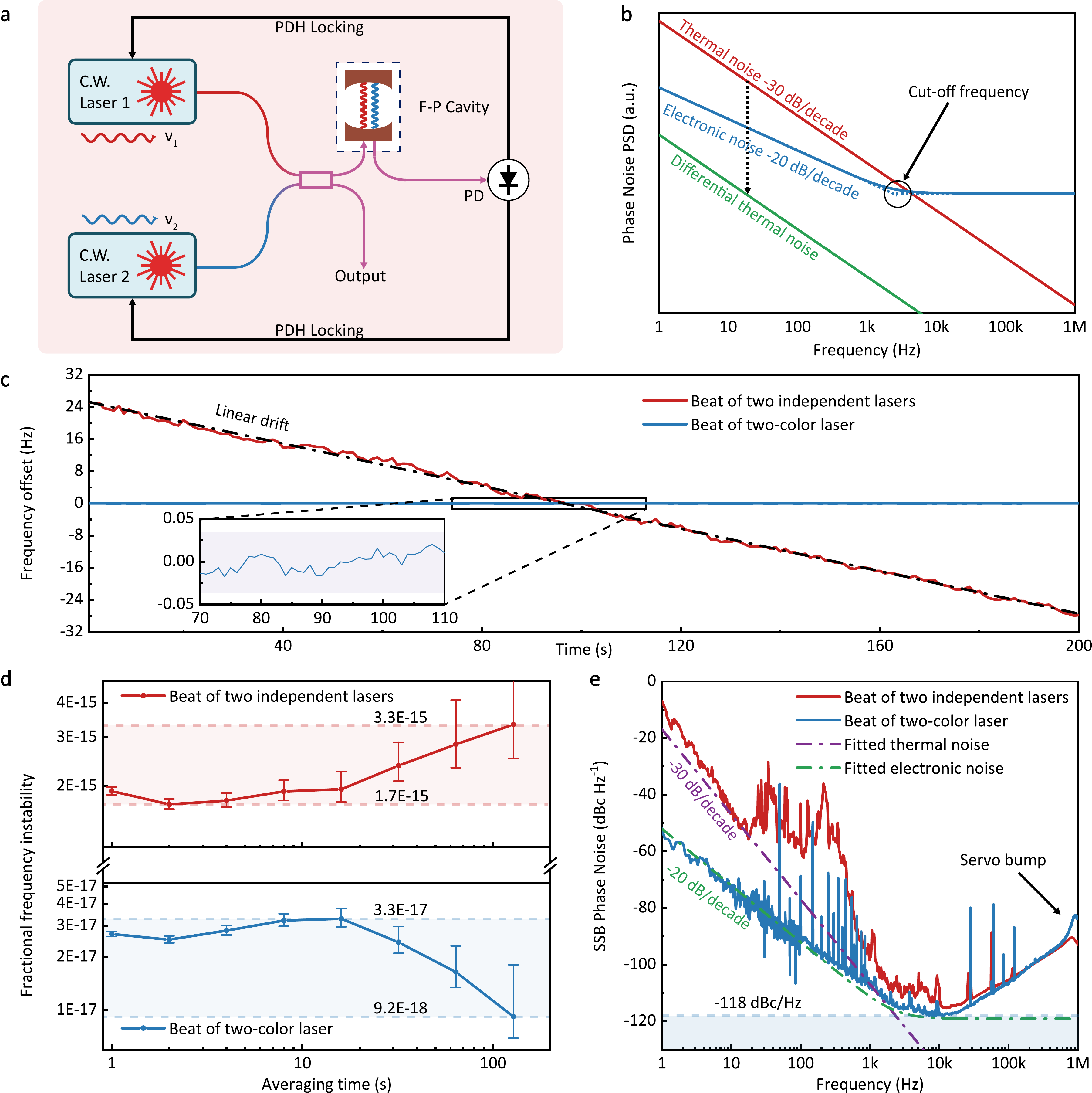}
\caption{\textbf{Characteristics of the two-color laser with a 1.5 GHz spacing.} 
\textbf{a,} Diagram of the two-color laser structure. Two independent continuous-wave lasers are phase-locked to a common reference cavity using the PDH method to create a two-color laser. 
\textbf{b,} Phase noise analysis of the two-color laser. 
\textbf{c,} Comparison of frequency fluctuations. The beat signal of two independent lasers displays a linear drift, whereas the beat signal of the two-color laser demonstrates significantly smaller fluctuations.
\textbf{d,} Comparison of fractional frequency instability. The fractional frequency instability of the beat signal of two independent lasers is predominantly influenced by thermal noise. In contrast, the beat signal of the two-color laser is primarily influenced by the electronic noise of the PDH locking, which is two orders of magnitude lower than that of the beat signal of two independent lasers.
\textbf{e,} Single-sideband (SSB) phase noise power spectral density analysis. The phase noise of the two-color laser's beat signal is markedly lower at lower Fourier frequencies compared to the beat signal of two independent lasers, underscoring the high coherence of the two-color laser. PDH Locking, Pound-Drever-Hall Locking; C.W., continuous-wave; PD, photodetector.
}
\label{fig2}
\end{figure*}

\vspace{6pt}
\noindent
\large\textbf{Results}\\
\noindent
\normalsize
\textbf{Principle of highly coherent two-color laser}\\
\noindent  
Figure \ref{fig2}a illustrates the schematic of the highly coherent two-color laser system. Two diode lasers are synchronized to a shared reference using the PDH method. Typically, a vacuum-mounted Fabry–Pérot cavity with exceptional finesse serves as the reference for PDH locking. The thermo-dynamically induced fluctuations in the optical path length of the cavity fundamentally limit the frequency stability of the locked laser.
When employing the PDH technique to stabilize a laser operating at a frequency $\nu$ within a cavity, the power spectral density (PSD) of its phase noise, denoted as $S_{\phi}^{\nu}\left( f \right)$, can be mathematically described as (see Supplementary Note 1 for more information)
\begin{equation}\label{eq1}
    \begin{aligned}
        S_{\phi}^{\nu}\left( f \right) =\frac{\nu ^2}{f^2}S_y\left( f \right) +S_e\left( f \right).
    \end{aligned}
\end{equation}
Here, $f$ represents the Fourier frequency offset, while $S_y$ signifies the thermo-dynamically optical length noise of the reference cavity, referred to as thermal noise in the subsequent discussion. $S_e$ denotes the electronic noise originating from the locking system. These distinct noise sources exert varying behaviors on the phase noise PSD curves, as depicted in Fig. \ref{fig2}b. The thermal noise manifests as a linear decrease with a slope of $-30$ dB/decade in the phase noise PSD (red curve). The blue curve in Fig. \ref{fig2}b delineates the electronic noise. It exhibits white phase noise beyond the cut-off frequency, equivalent to half the linewidth of the optical reference cavity. Below the cut-off frequency, it demonstrates a slope of $-20$ dB/decade. Typically, electronic noise is notably lower than thermal noise at lower Fourier frequencies. This observation prompts us to consider that if two diode lasers are synchronized to the same cavity, they will share correlated thermal noise. Within a specific frequency spacing, their relative phase noise will be constrained solely by electronic noise, potentially enabling the realization of a highly coherent two-color laser system.


In light of the phase noise characteristic of the PDH-locked laser, we can determine the relative phase noise of the two-color laser as (see Supplementary Note 1 for more information)
\begin{equation}\label{eq2}
    \begin{aligned}
       S_{\phi}^{\nu _2-v_1}\left( f \right) =\frac{\left( \nu _2-\nu _1 \right) ^2}{f^2}S_y\left( f \right) +S_{e_1}\left( f \right) +S_{e_2}\left( f \right). 
    \end{aligned}
\end{equation}
Interestingly, owing to the shared reference cavity between the two diode lasers, the thermo-dynamically length noise of the optical reference cavity exhibits a significant correlation. Its impact on the relative phase noise of the two-color laser is reduced by ${(\nu_2-\nu _1)}^2{/}{\nu_1}^2$ when compared to a one-color laser. We designate the initial term $S_{\rm{diff}}=\left( \nu_2-\nu_1 \right) ^2/f^2 \times S_y$ in Eq.\ref{eq2} as the differential thermal noise. As the frequency spacing $\nu_2-\nu_1$ expands, this differential thermal noise gradually increases. Nevertheless, within specific frequency spacing ranges, the contribution of this differential thermal noise remains well below that of the electronic noise, as is illustrated by the green curve in Fig. \ref{fig2}b. For the two-color laser, since two distinct servo systems are employed, the electronic noise of each system independently influences the relative phase noise of the two-color laser. Consequently, under these conditions, the relative phase noise of the two-color laser is predominantly dictated by the electronic noise of the two PDH locking systems, leading to exceptionally high coherence of two-color lasers.


\vspace{3pt}
\noindent 
\maketitle
\textbf{Two-color laser with a 1.5 GHz spacing }\\
\noindent To validate the coherence of the proposed two-color laser, we initially set the frequency spacing $\nu _2-\nu _1$ to 1.5 GHz, which is accomplished by synchronizing two diode lasers to a pair of adjacent modes of the optical reference cavity via the PDH method. The diode lasers operate near 1550 nm, while the optical cavity has a length of 10 cm and a finesse of approximately 330,000.
By doing so, the differential thermal noise is minimized in its contribution to the relative phase noise of the two-color laser. However, aligning the electrical noise PSD of the PDH locking mechanism with the quantum noise poses a notable challenge, particularly at lower frequency offsets. The presence of residual amplitude modulation (RAM) noise in the PDH locking, a consequence of imperfect phase modulation, significantly affects the electrical noise PSD at lower frequency offsets. To mitigate this effect, we implement several strategies. We utilize a low-RAM fiber-based phase modulator integrated with polarizers, enclosed in foam-filled aluminum enclosures to stabilize temperature fluctuations \cite{ZhangweiRAM, CLSRAM}. 
Besides, the parasitic étalons caused by the free space optical components also contribute to a RAM effect on the PDH locking system. To address this, we introduce slight tilts in all free-space optical components and employ isolators to prevent the formation of parasitic étalons \cite{shenhuiRAM}. Furthermore, we also use fiber devices with short pigtails to reduce the impact of non-common link sections in the two PDH locking loops. These optimizations bring the electrical noise near its quantum noise limitation. 

We demonstrate the high coherence of the two-color laser by comparing its stability and phase noise PSD with that of the one-color laser. In the case of the one-color laser, we establish two independent PDH systems, with each diode laser being locked to two independent optical reference cavity references. The beat note generated by these two independent lasers will reflect the stability and phase noise PSD properties of the one-color laser. Conversely, for the two-color laser, the beat note at 1.5 GHz will exhibit its distinctive characteristics. Figure \ref{fig2}c illustrates the comparison of frequency fluctuations. The beat of independent lasers demonstrates a linear drift over a short period, whereas the beat of the two-color laser exhibits significantly less fluctuation, indicating superior relative frequency stability. Stability analysis using fractional frequency instability is depicted in Fig. \ref{fig2}d. For the beat of two independent lasers, the fractional frequency instability is $1.9 \times 10^{-15}$ at 1 s, normalized to the optical carrier frequency. In contrast, the beat of the two-color laser shows a fractional frequency instability of $2.7 \times 10^{-17}$ at 1 s, which is two orders of magnitude lower, further highlighting the high relative stability of the two-color laser (see Supplementary Note 3 for additional details on the characterization).

Figure \ref{fig2}e illustrates the comparison of phase noise PSD between the two-color laser and the one-color laser. When considering the beat frequency of two independent lasers, the phase noise reaches $-$6 $\rm {dBc\ Hz^{-1}}$ at 1 Hz. It is important to note that apart from thermo-dynamically fluctuations in the reference cavity optical length, variations in residual gas pressure, as well as seismic or acoustic vibrations within the vacuum chamber, can also lead to fluctuations in the cavity optical length. These fluctuations manifest as laser frequency variations, contributing to the intricate phase noise observed in the beat of two independent lasers, as depicted in Fig. \ref{fig2}e. However, as per Eq.\ref{eq2}, these disturbances can be notably mitigated in the two-color laser system owing to the common-mode rejection. The SSB phase noise PSD of the beat frequency in the two-color laser system dips to $-$52 $\rm {dBc\ Hz^{-1}}$ at 1 Hz, $-$92 $\rm {dBc\ Hz^{-1}}$ at 100 Hz, exhibiting a slope of $-$20 dB/decade, and reaching a minimum of $-$118 $\rm {dBc\ Hz^{-1}}$ around a 10 kHz offset, which is close to the quantum noise of the photodetection in our scenario. This suggests that the predominant source of phase noise in the two-color laser primarily stems from the electronic noise present in the two PDH locking systems within their servo bandwidths.

\begin{figure*}[htbp]
\centering
\captionsetup{singlelinecheck=no, justification = RaggedRight}
\includegraphics[width=17.4cm]{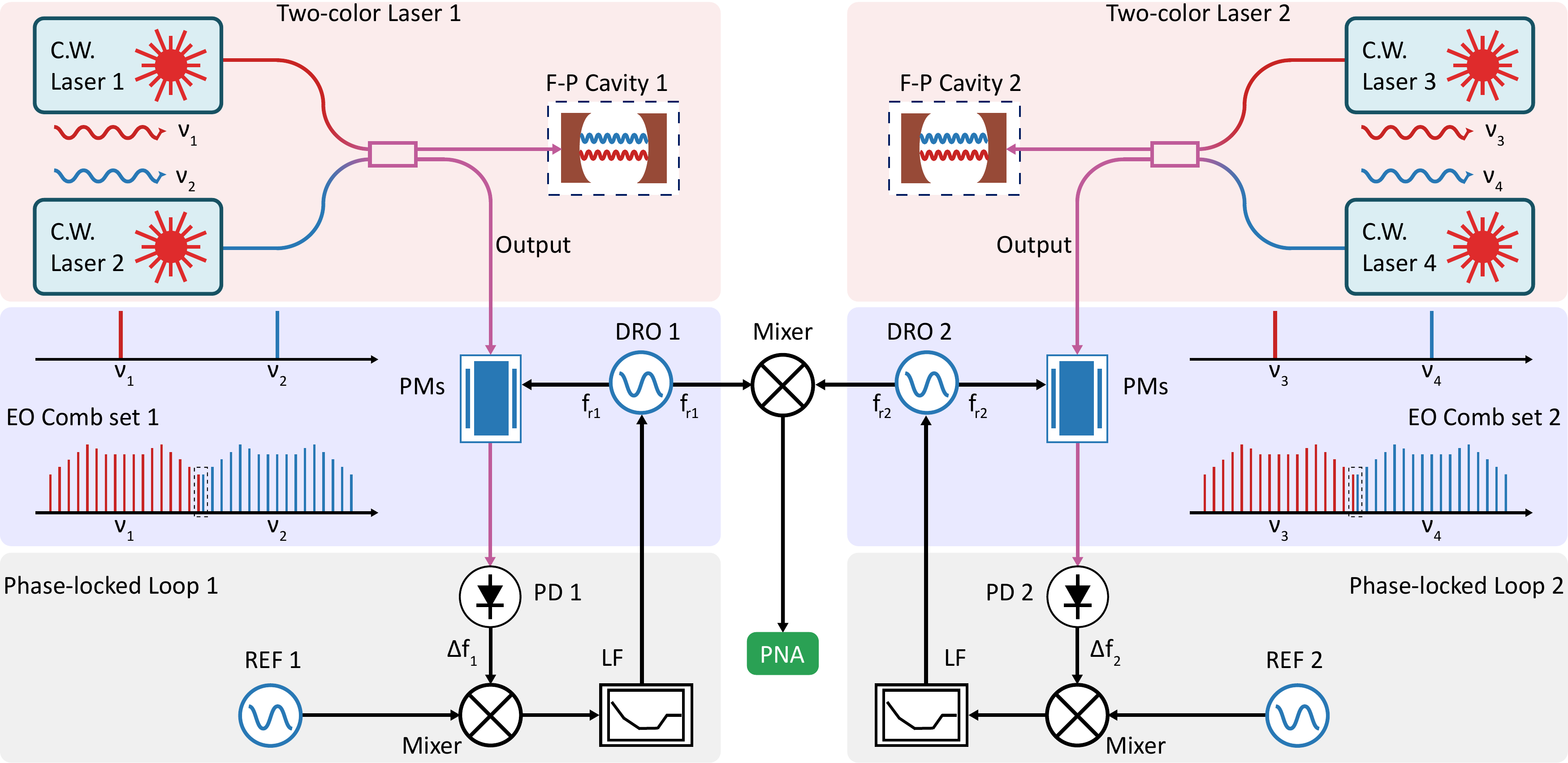}
\caption{\textbf{Phase noise characterization configuration for the two-color laser with 0.5 THz spacing.} We establish two distinct EO frequency comb-based optical frequency division systems. Within each system, the output of the two-color laser traverses phase modulators to generate two sets of EO frequency combs. These phase modulators are driven by a DRO. Comb teeth with overlapping spectra are filtered using a tunable optical filter and then directed to a photodetector to capture a beat signal. This beat signal is subsequently compared in-phase with a low-noise reference oscillator to generate the error signal that feeds back to the DRO. This configuration enables the transfer of the relative stability of the two-color laser to the DRO output signal. By beating the two DRO output signals, low carrier frequency signals are produced for assessing the relative phase noise and frequency stability of the two-color lasers. EO comb, Electro-Optical comb; PM, phase modulator; REF, reference oscillator; LF, loop filter; DRO, Dielectric resonator oscillator.
}
\label{fig3}
\end{figure*}

We emphasize that the fractional frequency instability achieved by the two-color laser surpasses that of previous reports \cite{Hall,chenqunfeng}. While a similar characterization approach was initially used in the early stages of introducing the PDH technique to assess its synchronization ability, this is the first time it has been demonstrated that the relative phase noise PSD of the two-color laser is solely constrained by the system's quantum noise, particularly at low-frequency offsets. Furthermore, as indicated by Eq.\ref{eq2}, increasing the frequency spacing of the two-color laser will not impact the SSB phase noise PSD of the beat within a certain range. This aspect, rarely discussed in previous studies, underscores the robustness of the two-color laser's phase noise performance.


\vspace{3pt}
\noindent 
\maketitle
\textbf{Two-color laser with a 0.5 THz spacing and its application for low noise microwave generation}\\
\noindent To further validate the advantages of the two-color laser, we increase its frequency spacing $\nu_2 - \nu_1$. According to Eq.\ref{eq2}, the differential thermal noise gradually increases as the frequency spacing of the two-color laser is widened. Based on the test results from the previous subsection, we anticipate that when the frequency spacing $\nu_2 - \nu_1$ of the two-color laser reaches 0.5 THz, the noise suppression factor will be $20\times\rm{log}(194~\text{THz}/0.5~\text{THz})$ = 51 dB. The thermal noise contribution is expected to reduce to $-$60 $\rm {dBc\ Hz^{-1}}$ at 1 Hz, approximately 10 dB lower than the electronic noise, suggesting that the relative stability of the two-color laser is still primarily influenced by electronic noise. However, as the frequency spacing $\nu_2 - \nu_1$ surpasses the bandwidth of the PD and testing equipment, characterizing the relative phase noise of the two-color laser poses a new challenge.

\begin{figure*}[htbp]
\centering
\captionsetup{singlelinecheck=no, justification = RaggedRight}
\includegraphics[width=17.4cm]{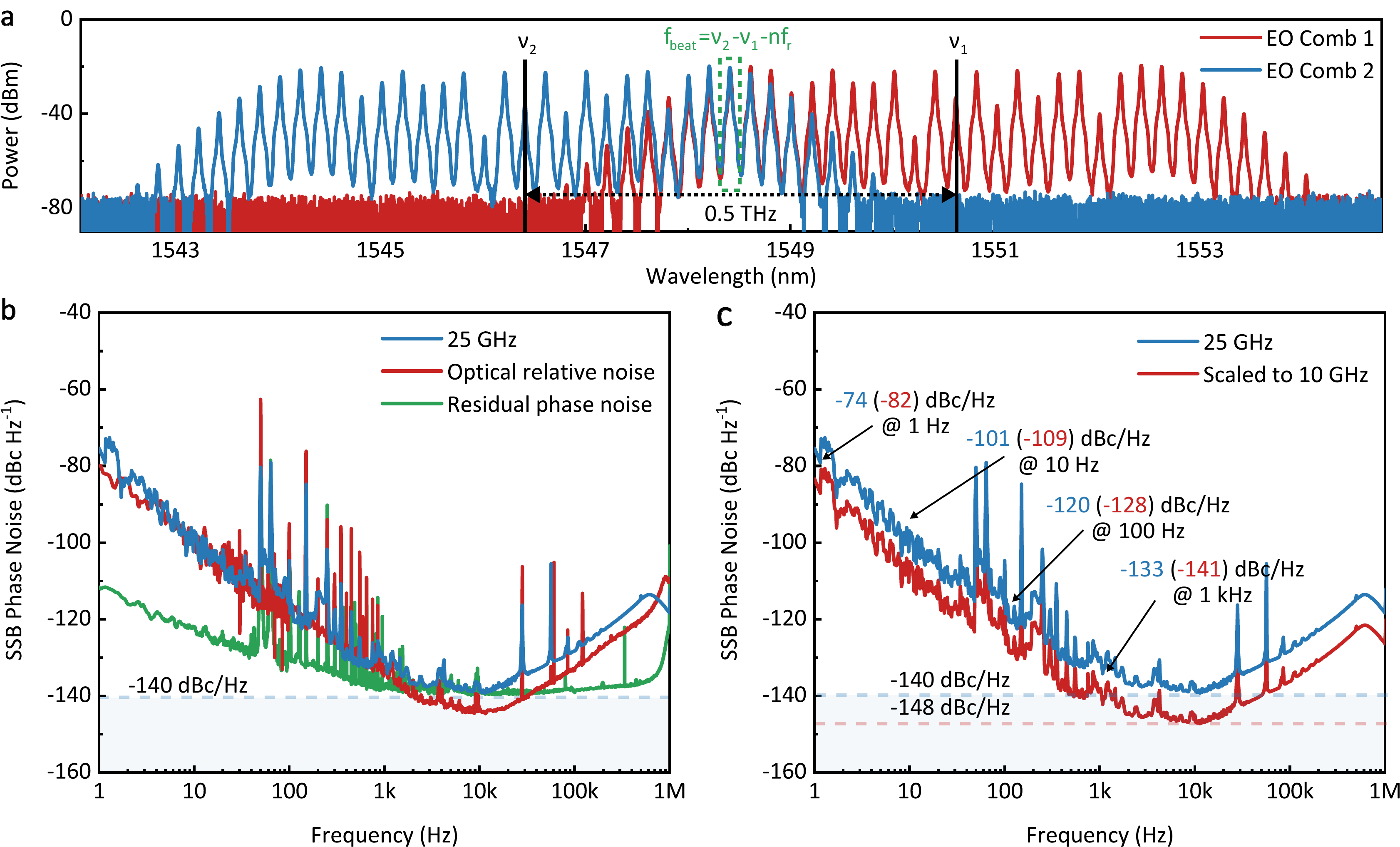}
\caption{\textbf{Experimental results of two-color laser with a 0.5 THz spacing and its application for low phase noise microwave generation.} \textbf{a,} EO frequency comb spectrum. The output of the two-color laser passes through phase modulators to generate two sets of EO frequency combs. \textbf{b,} Phase noise analysis of the 25 GHz signal generated through EO frequency division. At low Fourier frequency offsets, the phase noise of the generated 25 GHz signal closely aligns with the relative phase noise of the two-color laser when normalized to the microwave signal. However, at higher Fourier frequency offsets, the phase noise is constrained by the noise floor of the characterization system. Specifically, the residual phase noise of the EO frequency comb-based optical frequency division system. \textbf{c,} SSB phase noise of 25 GHz and its projection to 10 GHz. The phase noise characterization system produces an exceptional 25 GHz microwave signal.
}
\label{fig4}
\end{figure*}

Optical frequency combs, coherently bridging the optical and microwave domains, are expected to accurately assess the relative frequency stability and phase noise of a two-color laser with significant frequency spacings \cite{OFC1, OFC2}. A conventional femtosecond optical frequency comb can be utilized to transfer the stability of one frequency of the two-color laser to all comb teeth. By subsequently comparing another frequency of the two-color laser against the comb, one can obtain the relative SSB phase noise of the two-color laser \cite{transfer}. However, this approach is intricate and costly. Instead of this method, we opt for a simpler EO comb-based optical frequency division setup. Through EO frequency division, the relative stability of the two-color laser can be translated to a microwave signal, driving source of the EO frequency comb \cite{OFD_LJ, EOFC}. Following EO frequency division, the SSB phase noise of the resulting microwave signal will be $20 \times$ log $\left( \left( \nu _2-\nu _1 \right) /f_r \right) $ dB lower than the SSB phase noise of the two-color laser depicted in Fig. \ref{fig2}e. Here, $f_r$ denotes the repetition frequency of the EO frequency comb, corresponding to the driving signal frequency.
We anticipate that the SSB phase noise of the microwave signal will significantly undercut the sensitivity range of commercial phase noise analyzers. Thus, it is imperative to set up two similar configurations comprising two-color lasers and EO frequency division systems to produce two microwave signals with slightly different frequencies, subsequently combining them for characterization, as illustrated in Fig. \ref{fig3}. We emphasize that this scheme also offers a highly robust method for generating ultra-stable and low-phase noise microwave signals, which further demonstrates the considerable coherent advantages of the two-color laser.

Specifically, we characterize the relative phase noise of the two-color laser using the setup depicted in Fig. \ref{fig3}. The output of the two-color laser passes through two phase modulators driven by a dielectric resonator oscillator (DRO) operating at a frequency of $f_{r}$ to produce two sets of EO frequency comb. An optical filter selects comb teeth within the overlapping spectrum, and their beat frequency $f_{beat}=\nu _2-\nu _1-n\times f_r$ is detected on an amplified PD, where $n$ is the number of comb teeth between the two frequencies of the two-color laser. This beat frequency is then phase-compared with a low-noise reference frequency $f_{ref}$, and the resultant error signal is fed back to the 25 GHz DRO for phase locking. Upon achieving phase lock, the frequency $f_{r}$ is adjusted such that $f_{r}=(\nu _2-\nu _1-f_{ref})/n$, facilitating optical frequency division with $n$ representing the division factor.
The phase noise of the reference frequency $f_{ref}$ is much lower than the relative phase noise of the two-color laser. Therefore, the phase noise of $f_{r}$ is mainly determined by the relative phase noise of the two-color laser.
Considering the aforementioned analysis and the characteristics of the devices at our disposal, we set $\nu _2-\nu _1$ = 525 GHz, corresponding to a spacing of 21 comb teeth in the experimental setup. Consequently, the relative phase noise PSD of the two-color laser is reduced by $20\times\rm{log}(21)$ = 26 dB and transferred to the phase noise of the 25 GHz signal. The phase noise of the beat signals of the two sets of 25 GHz microwave signals reflects their absolute phase noise, enabling the inference of the relative phase noise of the two-color laser (see Supplementary Note 3 for more detail).

Figure \ref{fig4}a displays the optical spectrum of the EO frequency comb produced by one of the two-color lasers. In Fig. \ref{fig4}b, the blue curve illustrates the SSB phase noise PSD of the resulting 25 GHz signal. Additionally, we overlay the projection of the relative phase noise of the two-color laser, derived by shifting the SSB phase noise PSD of the two-color laser in Fig. \ref{fig2}e downward by 26 dB, for comparison within the same graph. At lower Fourier frequencies, the two curves exhibit a close alignment, maintaining a linear trend with a slope of $-$20 dB/decade. This suggests that the relative phase noise of the two-color laser remains influenced by the electronic noise of the PDH locking mechanism. Conversely, at higher Fourier frequencies, the phase noise of the down-converted 25 GHz signal is constrained by the residual phase noise of the optical frequency division system (depicted by the green curve in Fig. \ref{fig4}b). This residual phase noise is acquired through the division of the same two-color laser using two sets of EO frequency comb-based optical frequency division systems (see Supplementary Note 4 for additional details). It is noteworthy that this phase noise characterization system yields a high-performance 25 GHz signal with an SSB phase noise level of $-$74 dBc/Hz at 1 Hz. To provide a comparative analysis, this value is normalized to 10 GHz (red curve in Fig. \ref{fig4}c), showcasing the superior performance of the two-point frequency division method \cite{OFD_LJ,LJoptica,nature1,nature2,nature3,YQF,Vahala,LJarx,kudelin2024tunable,zhang2024monolithic,liu2024low,kittlaus2021low,Brillouin}, which achieves a fractional frequency instability of $6 \times 10^{-14}$ at 1 second (see Supplementary Note 3). This surpasses the short-term stability of hydrogen masers and most microwave atomic clocks \cite{Hmaser}.

It is noteworthy that the residual phase noise of the EO comb-based optical frequency division system achieves a level of $-$110 dBc/Hz at 1 Hz, as depicted in Fig. \ref{fig4}c. This level closely approaches that of the purest optical microwave signals currently accessible \cite{xie}, while the architecture of the frequency division system remains significantly simpler. The anticipated fractional instability at 1 second is approximately $2 \times 10^{-16}$, showcasing the considerable potential of the EO comb-based optical frequency division system in ultra-stable microwave signal generation. Compared with traditional optical frequency division methods \cite{xie,2011OFD}, leveraging the EO comb-based optical frequency division to translate the relative stability of a two-color laser into microwaves and produce low-noise microwave signals offers distinct advantages. Firstly, it eliminates the need to optimize various noise sources as required in traditional methods to constrain the laser's performance to thermal noise limitations, thereby enabling the generation of ultra-stable microwave signals with a more straightforward structure.
Secondly, this approach obviates the necessity for high-linearity PD and octave-spanning femtosecond optical frequency combs. The repetition rate of the EO frequency comb is directly determined by the driving frequency, eliminating the requirement for repetition rate multiplier devices such as Mach–Zehnder interferometers in photonics microwave generation. This method presents unique benefits in terms of cost-effectiveness, miniaturization, and portability \cite{xie, 2011OFD, NISTscience}.

\vspace{6pt}
\noindent
\large\textbf{Discussion} \\
\normalsize
\noindent  
In summary, we have made record-breaking progress in high-coherence two-color laser and ultra-stable microwave generation. The fractional frequency instability of the two-color laser has reached an impressive $2.7\times 10^{-17}$ at 1 s, normalized to the optical carrier frequency. Notably, the two-color laser maintains high coherence even with a frequency spacing of 0.5 THz. We characterized this relative stability using the EO frequency division, generating a 25 GHz microwave signal with a fractional frequency instability of $6\times 10^{-14}$ at 1 s. This achievement represents the most stable microwave signal produced through the two-point frequency division method. Furthermore, the residual fractional frequency instability of our EO frequency division system has reached $2\times 10^{-16}$ at 1 s, surpassing current benchmarks for stable photonics microwave generation and showcasing the strong potential of the EO frequency division system.

Further enhancements in the coherence and relative stability of the two-color laser are within reach. Opting for an optical reference cavity with a higher finesse can lower the cut-off frequency of the electronic noise PSD, thereby improving phase noise performance at lower Fourier frequency. 
Boosting the input optical power of the photoelectric detection system in PDH setups is expected to reduce quantum noise within PDH locking systems.
Conversely, actively managing RAM noise and the relative intensity noise of the seed lasers can significantly enhance both the short-term and long-term stability of the two-color laser and the generated microwave \cite{RIN}. Moreover, increasing the division factor could further enhance the stability of the generated microwave. Additionally, locking to an atomic reference line can reinforce the long-term stability of the two-color laser and the generated microwave \cite{NISTscience}.
To diminish the residual phase noise floor of the EO frequency division system, boosting the signal-to-noise ratio of the EO frequency comb with a low half-wave voltage phase modulator and employing a high-frequency, high-power photodetector for direct detection prove to be effective strategies \cite{kudelin2024tunable, modulator1, modulator2}. These interventions are poised to elevate the phase noise performance of the generated microwave signal at higher Fourier frequency.

The high-coherence two-color laser and ultra-stable low-noise microwave signal showcased in this study are poised to play pivotal roles across various applications. This remarkably coherent two-color laser is anticipated to enhance measurement precision in high-precision optical interferometers, like those utilized in gravitational wave detection. Moreover, it is expected to be instrumental in precision spectroscopy, encompassing dual-comb systems and biological imaging. In realms involving precise light-matter interactions like CPT atomic clocks, entangled quantum state preparation, quantum information, and quantum computing, this highly coherent two-color laser stands to make a substantial impact. Low-noise terahertz signals can be generated by directly converting the two-color laser with a high-speed photodiode.
Undoubtedly, the high-stability and low-noise microwave signal generated in this work hold significant value for applications in radar \cite{app2}, navigation, very long-baseline interferometry, deep space exploration, and time-frequency standards such as microwave atomic clocks \cite{app1}.

\vspace{6pt}
\begin{footnotesize}

\vspace{6pt}
\noindent\textbf{Data availability}\\
The data that supports the plots within this paper and other findings of this study are available from the corresponding authors upon request.

\vspace{6pt}
\noindent\textbf{Code availability}\\
The codes that support the findings of this study are available from the corresponding authors upon request.

\end{footnotesize}
\vspace{20pt}

\bibliography{main.bib}

\begin{thebibliography}{10}
\expandafter\ifx\csname url\endcsname\relax
  \def\url#1{\texttt{#1}}\fi
\expandafter\ifx\csname urlprefix\endcsname\relax\def\urlprefix{URL }\fi
\providecommand{\bibinfo}[2]{#2}
\providecommand{\eprint}[2][]{\url{#2}}

\bibitem{1gravitational}
\bibinfo{author}{Abramovici, A.} \emph{et~al.}
\newblock \bibinfo{title}{\href{https://doi.org/10.1126/science.256.5055.325}{LIGO: The laser interferometer gravitational-wave observatory}}.
\newblock \emph{\bibinfo{journal}{Science}} \textbf{\bibinfo{volume}{256}}, \bibinfo{pages}{325--333} (\bibinfo{year}{1992}).

\bibitem{2gravitational}
\bibinfo{author}{Abbott, B.~P.} \emph{et~al.}
\newblock \bibinfo{title}{\href{https://doi.org/10.1103/PhysRevLett.116.061102}{Observation of gravitational waves from a binary black hole merger}}.
\newblock \emph{\bibinfo{journal}{Physical Review Letters}} \textbf{\bibinfo{volume}{116}}, \bibinfo{pages}{061102} (\bibinfo{year}{2016}).

\bibitem{interferometry}
\bibinfo{author}{Ishii, Y.} \& \bibinfo{author}{Onodera, R.}
\newblock \bibinfo{title}{\href{https://doi.org/10.1364/OL.16.001523}{Two-wavelength laser-diode interferometry that uses phase-shifting techniques}}.
\newblock \emph{\bibinfo{journal}{Optics Letters}} \textbf{\bibinfo{volume}{16}}, \bibinfo{pages}{1523--1525} (\bibinfo{year}{1991}).

\bibitem{Dual-comb}
\bibinfo{author}{Coddington, I.}, \bibinfo{author}{Newbury, N.} \& \bibinfo{author}{Swann, W.}
\newblock \bibinfo{title}{\href{https://doi.org/10.1364/OPTICA.3.000414}{Dual-comb spectroscopy}}.
\newblock \emph{\bibinfo{journal}{Optica}} \textbf{\bibinfo{volume}{3}}, \bibinfo{pages}{414--426} (\bibinfo{year}{2016}).

\bibitem{tsao2024dual}
\bibinfo{author}{Tsao, E.~J.} \emph{et~al.}
\newblock \bibinfo{title}{\href{https://doi.org/10.48550/arXiv.2405.14842}{Dual-comb correlation spectroscopy of thermal light}}.
\newblock \emph{\bibinfo{journal}{arXiv preprint arXiv:2405.14842}}  (\bibinfo{year}{2024}).

\bibitem{picque2019frequency}
\bibinfo{author}{Picqu{\'e}, N.} \& \bibinfo{author}{H{\"a}nsch, T.~W.}
\newblock \bibinfo{title}{\href{https://doi.org/10.1038/s41566-018-0347-5}{Frequency comb spectroscopy}}.
\newblock \emph{\bibinfo{journal}{Nature Photonics}} \textbf{\bibinfo{volume}{13}}, \bibinfo{pages}{146--157} (\bibinfo{year}{2019}).

\bibitem{2CPT}
\bibinfo{author}{Bergmann, K.}, \bibinfo{author}{Theuer, H.} \& \bibinfo{author}{Shore, B.}
\newblock \bibinfo{title}{\href{https://doi.org/10.1103/RevModPhys.70.1003}{Coherent population transfer among quantum states of atoms and molecules}}.
\newblock \emph{\bibinfo{journal}{Reviews of Modern Physics}} \textbf{\bibinfo{volume}{70}}, \bibinfo{pages}{1003} (\bibinfo{year}{1998}).

\bibitem{EIT}
\bibinfo{author}{Fleischhauer, M.}, \bibinfo{author}{Imamoglu, A.} \& \bibinfo{author}{Marangos, J.~P.}
\newblock \bibinfo{title}{\href{https://doi.org/10.1103/RevModPhys.77.633}{Electromagnetically induced transparency: Optics in coherent media}}.
\newblock \emph{\bibinfo{journal}{Reviews of Modern Physics}} \textbf{\bibinfo{volume}{77}}, \bibinfo{pages}{633--673} (\bibinfo{year}{2005}).

\bibitem{quantum}
\bibinfo{author}{Madsen, L.~S.} \emph{et~al.}
\newblock \bibinfo{title}{\href{https://doi.org/10.1038/s41586-022-04725-x}{Quantum computational advantage with a programmable photonic processor}}.
\newblock \emph{\bibinfo{journal}{Nature}} \textbf{\bibinfo{volume}{606}}, \bibinfo{pages}{75--81} (\bibinfo{year}{2022}).

\bibitem{OFD_LJ}
\bibinfo{author}{Li, J.}, \bibinfo{author}{Yi, X.}, \bibinfo{author}{Lee, H.}, \bibinfo{author}{Diddams, S.~A.} \& \bibinfo{author}{Vahala, K.~J.}
\newblock \bibinfo{title}{\href{https://doi.org/10.1126/science.1252909}{Electro-optical frequency division and stable microwave synthesis}}.
\newblock \emph{\bibinfo{journal}{Science}} \textbf{\bibinfo{volume}{345}}, \bibinfo{pages}{309--313} (\bibinfo{year}{2014}).

\bibitem{LJoptica}
\bibinfo{author}{Li, J.} \& \bibinfo{author}{Vahala, K.}
\newblock \bibinfo{title}{\href{https://doi.org/10.1364/OPTICA.477602}{Small-sized, ultra-low phase noise photonic microwave oscillators at X-Ka bands}}.
\newblock \emph{\bibinfo{journal}{Optica}} \textbf{\bibinfo{volume}{10}}, \bibinfo{pages}{33--34} (\bibinfo{year}{2023}).

\bibitem{nature1}
\bibinfo{author}{Sun, S.} \emph{et~al.}
\newblock \bibinfo{title}{\href{https://doi.org/10.1038/s41586-024-07057-0}{Integrated optical frequency division for microwave and mmWave generation}}.
\newblock \emph{\bibinfo{journal}{Nature}} \textbf{\bibinfo{volume}{627}}, \bibinfo{pages}{540--545} (\bibinfo{year}{2024}).

\bibitem{nature2}
\bibinfo{author}{Kudelin, I.} \emph{et~al.}
\newblock \bibinfo{title}{\href{https://doi.org/10.1038/s41586-024-07058-z}{Photonic chip-based low-noise microwave oscillator}}.
\newblock \emph{\bibinfo{journal}{Nature}} \textbf{\bibinfo{volume}{627}}, \bibinfo{pages}{534--539} (\bibinfo{year}{2024}).

\bibitem{nature3}
\bibinfo{author}{Zhao, Y.} \emph{et~al.}
\newblock \bibinfo{title}{\href{https://doi.org/10.1038/s41586-024-07136-2}{All-optical frequency division on-chip using a single laser}}.
\newblock \emph{\bibinfo{journal}{Nature}} \textbf{\bibinfo{volume}{627}}, \bibinfo{pages}{546--552} (\bibinfo{year}{2024}).

\bibitem{YQF}
\bibinfo{author}{Jin, X.} \emph{et~al.}
\newblock \bibinfo{title}{\href{https://doi.org/10.48550/arXiv.2401.12760}{Microresonator-referenced soliton microcombs with zeptosecond-level timing noise}}.
\newblock \emph{\bibinfo{journal}{arXiv preprint arXiv:2401.12760}}  (\bibinfo{year}{2024}).

\bibitem{Vahala}
\bibinfo{author}{Ji, Q.-X.} \emph{et~al.}
\newblock \bibinfo{title}{\href{https://doi.org/10.48550/arXiv.2403.00973}{Dispersive-wave-agile optical frequency division}}.
\newblock \emph{\bibinfo{journal}{arXiv preprint arXiv:2403.00973}}  (\bibinfo{year}{2024}).

\bibitem{LJarx}
\bibinfo{author}{He, Y.} \emph{et~al.}
\newblock \bibinfo{title}{\href{https://doi.org/10.1126/sciadv.ado9570}{Chip-scale high-performance photonic microwave oscillator}}.
\newblock \emph{\bibinfo{journal}{Science Advances}} \textbf{\bibinfo{volume}{10}}, \bibinfo{pages}{eado9570} (\bibinfo{year}{2024}).

\bibitem{kudelin2024tunable}
\bibinfo{author}{Kudelin, I.} \emph{et~al.}
\newblock \bibinfo{title}{\href{https://doi.org/10.48550/arXiv.2404.00136}{Tunable X-band opto-electronic synthesizer with ultralow phase noise}}.
\newblock \emph{\bibinfo{journal}{arXiv preprint arXiv:2404.00136}}  (\bibinfo{year}{2024}).

\bibitem{zhang2024monolithic}
\bibinfo{author}{Zhang, W.} \emph{et~al.}
\newblock \bibinfo{title}{\href{https://doi.org/10.1038/s42005-024-01660-3}{Monolithic optical resonator for ultrastable laser and photonic millimeter-wave synthesis}}.
\newblock \emph{\bibinfo{journal}{Communications Physics}} \textbf{\bibinfo{volume}{7}}, \bibinfo{pages}{177} (\bibinfo{year}{2024}).

\bibitem{liu2024low}
\bibinfo{author}{Liu, Y.} \emph{et~al.}
\newblock \bibinfo{title}{\href{https://doi.org/10.1063/5.0174544}{Low-noise microwave generation with an air-gap optical reference cavity}}.
\newblock \emph{\bibinfo{journal}{APL Photonics}} \textbf{\bibinfo{volume}{9}} (\bibinfo{year}{2024}).

\bibitem{kittlaus2021low}
\bibinfo{author}{Kittlaus, E.~A.} \emph{et~al.}
\newblock \bibinfo{title}{\href{https://doi.org/10.1038/s41467-021-24637-0}{A low-noise photonic heterodyne synthesizer and its application to millimeter-wave radar}}.
\newblock \emph{\bibinfo{journal}{Nature Communications}} \textbf{\bibinfo{volume}{12}}, \bibinfo{pages}{4397} (\bibinfo{year}{2021}).

\bibitem{Brillouin}
\bibinfo{author}{Tetsumoto, T.} \emph{et~al.}
\newblock \bibinfo{title}{\href{https://doi.org/10.1038/s41566-021-00790-2}{Optically referenced 300 GHz millimetre-wave oscillator}}.
\newblock \emph{\bibinfo{journal}{Nature Photonics}} \textbf{\bibinfo{volume}{15}}, \bibinfo{pages}{516--522} (\bibinfo{year}{2021}).

\bibitem{two-mode}
\bibinfo{author}{Zhang, P.-J.} \emph{et~al.}
\newblock \bibinfo{title}{\href{https://doi.org/10.1073/pnas.2101605118}{Single-mode characteristic of a supermode microcavity Raman laser}}.
\newblock \emph{\bibinfo{journal}{Proceedings of the National Academy of Sciences}} \textbf{\bibinfo{volume}{118}}, \bibinfo{pages}{e2101605118} (\bibinfo{year}{2021}).

\bibitem{EOmod}
\bibinfo{author}{Wang, C.} \emph{et~al.}
\newblock \bibinfo{title}{\href{https://doi.org/10.1038/s41586-018-0551-y}{Integrated lithium niobate electro-optic modulators operating at CMOS-compatible voltages}}.
\newblock \emph{\bibinfo{journal}{Nature}} \textbf{\bibinfo{volume}{562}}, \bibinfo{pages}{101--104} (\bibinfo{year}{2018}).

\bibitem{PDH_Hall}
\bibinfo{author}{Drever, R.~W.} \emph{et~al.}
\newblock \bibinfo{title}{\href{https://doi.org/10.1007/BF00702605}{Laser phase and frequency stabilization using an optical resonator}}.
\newblock \emph{\bibinfo{journal}{Applied Physics B}} \textbf{\bibinfo{volume}{31}}, \bibinfo{pages}{97--105} (\bibinfo{year}{1983}).

\bibitem{40mHz}
\bibinfo{author}{Kessler, T.} \emph{et~al.}
\newblock \bibinfo{title}{\href{https://doi.org/10.1038/nphoton.2012.217}{A sub-40-mHz-linewidth laser based on a silicon single-crystal optical cavity}}.
\newblock \emph{\bibinfo{journal}{Nature Photonics}} \textbf{\bibinfo{volume}{6}}, \bibinfo{pages}{687--692} (\bibinfo{year}{2012}).

\bibitem{mHz}
\bibinfo{author}{Matei, D.} \emph{et~al.}
\newblock \bibinfo{title}{\href{https://doi.org/10.1103/PhysRevLett.118.263202}{1.5 $\upmu$m lasers with sub-10 mHz linewidth}}.
\newblock \emph{\bibinfo{journal}{Physical Review Letters}} \textbf{\bibinfo{volume}{118}}, \bibinfo{pages}{263202} (\bibinfo{year}{2017}).

\bibitem{clock}
\bibinfo{author}{Bloom, B.} \emph{et~al.}
\newblock \bibinfo{title}{\href{https://doi.org/10.1038/nature12941}{An optical lattice clock with accuracy and stability at the $10^{-18}$ level}}.
\newblock \emph{\bibinfo{journal}{Nature}} \textbf{\bibinfo{volume}{506}}, \bibinfo{pages}{71--75} (\bibinfo{year}{2014}).

\bibitem{xie}
\bibinfo{author}{Xie, X.} \emph{et~al.}
\newblock \bibinfo{title}{\href{https://doi.org/10.1038/nphoton.2016.215}{Photonic microwave signals with zeptosecond-level absolute timing noise}}.
\newblock \emph{\bibinfo{journal}{Nature Photonics}} \textbf{\bibinfo{volume}{11}}, \bibinfo{pages}{44--47} (\bibinfo{year}{2017}).

\bibitem{2011OFD}
\bibinfo{author}{Fortier, T.~M.} \emph{et~al.}
\newblock \bibinfo{title}{\href{https://doi.org/10.1038/nphoton.2011.121}{Generation of ultrastable microwaves via optical frequency division}}.
\newblock \emph{\bibinfo{journal}{Nature Photonics}} \textbf{\bibinfo{volume}{5}}, \bibinfo{pages}{425--429} (\bibinfo{year}{2011}).

\bibitem{NISTscience}
\bibinfo{author}{Nakamura, T.} \emph{et~al.}
\newblock \bibinfo{title}{\href{https://doi.org/10.1126/science.abb2473}{Coherent optical clock down-conversion for microwave frequencies with $10^{-18}$ instability}}.
\newblock \emph{\bibinfo{journal}{Science}} \textbf{\bibinfo{volume}{368}}, \bibinfo{pages}{889--892} (\bibinfo{year}{2020}).

\bibitem{yujialiang}
\bibinfo{author}{Yu, J.} \emph{et~al.}
\newblock \bibinfo{title}{\href{https://doi.org/10.1103/PhysRevX.13.041002}{Excess noise and photoinduced effects in highly reflective crystalline mirror coatings}}.
\newblock \emph{\bibinfo{journal}{Physical Review X}} \textbf{\bibinfo{volume}{13}}, \bibinfo{pages}{041002} (\bibinfo{year}{2023}).

\bibitem{Hall}
\bibinfo{author}{Salomon, C.}, \bibinfo{author}{Hils, D.} \& \bibinfo{author}{Hall, J.}
\newblock \bibinfo{title}{\href{https://doi.org/10.1364/JOSAB.5.001576}{Laser stabilization at the millihertz level}}.
\newblock \emph{\bibinfo{journal}{JOSA B}} \textbf{\bibinfo{volume}{5}}, \bibinfo{pages}{1576--1587} (\bibinfo{year}{1988}).

\bibitem{chenqunfeng}
\bibinfo{author}{Chen, Q.-F.}, \bibinfo{author}{Nevsky, A.} \& \bibinfo{author}{Schiller, S.}
\newblock \bibinfo{title}{\href{https://doi.org/10.1007/s00340-012-5014-9}{Locking the frequency of lasers to an optical cavity at the $1.6\times 10^{-17}$ relative instability level}}.
\newblock \emph{\bibinfo{journal}{Applied Physics B}} \textbf{\bibinfo{volume}{107}}, \bibinfo{pages}{679--683} (\bibinfo{year}{2012}).

\bibitem{ZhangweiRAM}
\bibinfo{author}{Zhang, W.} \emph{et~al.}
\newblock \bibinfo{title}{\href{https://doi.org/10.1364/OL.39.001980}{Reduction of residual amplitude modulation to $1\times 10^{-6}$ for frequency modulation and laser stabilization}}.
\newblock \emph{\bibinfo{journal}{Optics Letters}} \textbf{\bibinfo{volume}{39}}, \bibinfo{pages}{1980--1983} (\bibinfo{year}{2014}).

\bibitem{CLSRAM}
\bibinfo{author}{Li, L.}, \bibinfo{author}{Liu, F.}, \bibinfo{author}{Wang, C.} \& \bibinfo{author}{Chen, L.}
\newblock \bibinfo{title}{\href{https://doi.org/10.1063/1.4704084}{Measurement and control of residual amplitude modulation in optical phase modulation}}.
\newblock \emph{\bibinfo{journal}{Review of Scientific Instruments}} \textbf{\bibinfo{volume}{83}} (\bibinfo{year}{2012}).

\bibitem{shenhuiRAM}
\bibinfo{author}{Shen, H.}, \bibinfo{author}{Li, L.}, \bibinfo{author}{Bi, J.}, \bibinfo{author}{Wang, J.} \& \bibinfo{author}{Chen, L.}
\newblock \bibinfo{title}{\href{https://doi.org/10.1103/PhysRevA.92.063809}{Systematic and quantitative analysis of residual amplitude modulation in Pound-Drever-Hall frequency stabilization}}.
\newblock \emph{\bibinfo{journal}{Physical Review A}} \textbf{\bibinfo{volume}{92}}, \bibinfo{pages}{063809} (\bibinfo{year}{2015}).

\bibitem{OFC1}
\bibinfo{author}{H{\"a}nsch, T.~W.}
\newblock \bibinfo{title}{\href{https://doi.org/10.1103/RevModPhys.78.1297}{Nobel lecture: passion for precision}}.
\newblock \emph{\bibinfo{journal}{Reviews of Modern Physics}} \textbf{\bibinfo{volume}{78}}, \bibinfo{pages}{1297} (\bibinfo{year}{2006}).

\bibitem{OFC2}
\bibinfo{author}{Hall, J.~L.}
\newblock \bibinfo{title}{\href{https://doi.org/10.1103/RevModPhys.78.1279}{Nobel Lecture: Defining and measuring optical frequencies}}.
\newblock \emph{\bibinfo{journal}{Reviews of Modern Physics}} \textbf{\bibinfo{volume}{78}}, \bibinfo{pages}{1279} (\bibinfo{year}{2006}).

\bibitem{transfer}
\bibinfo{author}{Nicolodi, D.} \emph{et~al.}
\newblock \bibinfo{title}{\href{https://doi.org/10.1038/nphoton.2013.361}{Spectral purity transfer between optical wavelengths at the $10^{-18}$ level}}.
\newblock \emph{\bibinfo{journal}{Nature Photonics}} \textbf{\bibinfo{volume}{8}}, \bibinfo{pages}{219--223} (\bibinfo{year}{2014}).

\bibitem{EOFC}
\bibinfo{author}{Wu, R.}, \bibinfo{author}{Supradeepa, V.}, \bibinfo{author}{Long, C.~M.}, \bibinfo{author}{Leaird, D.~E.} \& \bibinfo{author}{Weiner, A.~M.}
\newblock \bibinfo{title}{\href{https://doi.org/10.1364/OL.35.003234}{Generation of very flat optical frequency combs from continuous-wave lasers using cascaded intensity and phase modulators driven by tailored radio frequency waveforms}}.
\newblock \emph{\bibinfo{journal}{Optics Letters}} \textbf{\bibinfo{volume}{35}}, \bibinfo{pages}{3234--3236} (\bibinfo{year}{2010}).

\bibitem{Hmaser}
\bibinfo{author}{Marlow, B. L.~S.} \& \bibinfo{author}{Scherer, D.~R.}
\newblock \bibinfo{title}{\href{https://doi.org/10.1109/TUFFC.2021.3049713}{A review of commercial and emerging atomic frequency standards}}.
\newblock \emph{\bibinfo{journal}{IEEE Transactions on Ultrasonics, Ferroelectrics, and Frequency Control}} \textbf{\bibinfo{volume}{68}}, \bibinfo{pages}{2007--2022} (\bibinfo{year}{2021}).

\bibitem{RIN}
\bibinfo{author}{Tricot, F.}, \bibinfo{author}{Phung, D.}, \bibinfo{author}{Lours, M.}, \bibinfo{author}{Gu{\'e}randel, S.} \& \bibinfo{author}{De~Clercq, E.}
\newblock \bibinfo{title}{\href{https://doi.org/10.1063/1.5046852}{Power stabilization of a diode laser with an acousto-optic modulator}}.
\newblock \emph{\bibinfo{journal}{Review of Scientific Instruments}} \textbf{\bibinfo{volume}{89}} (\bibinfo{year}{2018}).

\bibitem{modulator1}
\bibinfo{author}{Kharel, P.}, \bibinfo{author}{Reimer, C.}, \bibinfo{author}{Luke, K.}, \bibinfo{author}{He, L.} \& \bibinfo{author}{Zhang, M.}
\newblock \bibinfo{title}{\href{https://doi.org/10.1364/OPTICA.440484}{Breaking voltage--bandwidth limits in integrated lithium niobate modulators using micro-structured electrodes}}.
\newblock \emph{\bibinfo{journal}{Optica}} \textbf{\bibinfo{volume}{8}}, \bibinfo{pages}{357--363} (\bibinfo{year}{2021}).

\bibitem{modulator2}
\bibinfo{author}{Valdez, F.}, \bibinfo{author}{Mere, V.} \& \bibinfo{author}{Mookherjea, S.}
\newblock \bibinfo{title}{\href{https://doi.org/10.1364/OPTICA.484549}{100 GHz bandwidth, 1 volt integrated electro-optic Mach--Zehnder modulator at near-IR wavelengths}}.
\newblock \emph{\bibinfo{journal}{Optica}} \textbf{\bibinfo{volume}{10}}, \bibinfo{pages}{578--584} (\bibinfo{year}{2023}).

\bibitem{app2}
\bibinfo{author}{Soumya, A.}, \bibinfo{author}{Krishna~Mohan, C.} \& \bibinfo{author}{Cenkeramaddi, L.~R.}
\newblock \bibinfo{title}{\href{https://doi.org/10.3390/s23218901}{Recent advances in mmWave-radar-based sensing, its applications, and machine learning techniques: A review}}.
\newblock \emph{\bibinfo{journal}{Sensors}} \textbf{\bibinfo{volume}{23}}, \bibinfo{pages}{8901} (\bibinfo{year}{2023}).

\bibitem{app1}
\bibinfo{author}{Vig, J.~R.}
\newblock \bibinfo{title}{\href{https://doi.org/10.1109/58.238104}{Military applications of high accuracy frequency standards and clocks}}.
\newblock \emph{\bibinfo{journal}{IEEE Transactions on Ultrasonics, Ferroelectrics, and Frequency Control}} \textbf{\bibinfo{volume}{40}}, \bibinfo{pages}{522--527} (\bibinfo{year}{1993}).

\end{thebibliography}


\vspace{12pt}
\begin{footnotesize}

\vspace{6pt}
\noindent \textbf{Acknowledgment}

\noindent 
This work is supported by the Beijing Natural Science Foundation(grant no.JQ24027 to X.X.) and the National Natural Science Foundation of China (grant no.62071010 to X.X.). The authors acknowledge fruitful discussions with Dr. Xiaogang Zhang for the PDH locking.
The authors acknowledge Jingjing Lin for her assistance in polishing the English of the supplementary notes.

\vspace{6pt}
\noindent \textbf{Author contributions}

\noindent 
B.H., Jiachuan Yang, F.M., and X.X. jointly designed the experimental setup. F.M. and X.X. provided the initial basis for the two-color laser. B.H., Jiachuan Yang, F.M., C.Z., and X.X. analyzed and optimized two-color lasers. Jialiang Yu, N.Z., Y.L., and Z.F. contributed to the setup of the two-color lasers. Q.Y. provided one of the C.W. lasers. B.H. and Jiachuan Yang constructed the EO frequency comb and division system. Z.C. provided insights during noise analysis. B.H. and Jiachuan Yang conducted the experiments and collected the data. B.H., Jiachuan Yang, and X.X. wrote the manuscript. All co-authors provided valuable feedback and comments. X.X. led the program and supervised the work.
\vspace{6pt}

\noindent\textbf{Additional information} 

\noindent Supplementary information is available in the online version of the paper. Reprints and permissions information is available online. Correspondence and requests for materials should be addressed to F.M. and X.X.

\vspace{6pt}
\noindent \textbf{Competing financial interests} 

\noindent The authors declare no competing financial interests.
\end{footnotesize}

\end{document}